# Thermally Activated Motion of Dislocations in Fields of Obstacles: Effect of Obstacle Distribution

Zhijie Xu and R.C. Picu[1]

Department of Mechanical, Aerospace and Nuclear Engineering,

Rensselaer Polytechnic Institute, Troy, NY 12180

**Abstract**

The thermally activated motion of dislocations across fields of obstacles distributed at random and in a correlated manner, in separate models, is studied by means of computer simulations. The strain rate sensitivity and strength are evaluated in terms of the obstacle strength, temperature and applied shear stress. Above a threshold stress, the dislocation motion undergoes a transition from smooth to jerky, i.e. obstacles are overcome in a correlated manner at high stresses, while at low stresses they are overcome individually. This leads to a significant reduction of the strain rate sensitivity. The threshold stress depends on the obstacle strength and temperature. A model is proposed to predict the strain rate sensitivity and the smooth-to-jerky transition stress. Obstacle clustering has little effect on strain rate sensitivity at low stress (creep conditions), but it becomes important at high stress. It is concluded that models for the strength and strain rate sensitivity should include higher moments of the obstacle density distribution in addition to the mean density.

---

[1] Corresponding author: E-mail; picuc@rpi.edu

1. **Introduction**

The motion of dislocations across fields of obstacles was studied extensively in the past in connection with solid solution hardening. The first studies were dedicated to the prediction of the critical resolved shear stress for a population of identical obstacles randomly distributed in the glide plane. The classical result for this problem is due to Friedel [1] who showed that the critical stress scales with the obstacle strength, $f_c$, to power 3/2, while the mean spacing between obstacles in contact with the dislocation line scales with the applied stress to the -1/3 power. The scaling was confirmed by models developed by Morris and Klahn [2], Hanson and Morris [3], Labusch [4] and others, and by computer simulations [5,6,7]. These analytic models were developed under rather restrictive assumptions of no temporal and spatial correlations of obstacle failure events for given dislocation. Early models also required that the mean angle made by the dislocation line with the glide direction is 90º on the scale of the mean obstacle spacing. This restriction was relaxed by Labusch [4]; this led to a significant improvement of the agreement between the analytic and numerical results.

It has been suggested that these assumptions and models are adequate for distributions of weak obstacles which are overcome mostly by the "unzipping" mechanism. In this context, unzipping refers to the mechanism by which bypassing an obstacle of a stable dislocation configuration leads to bypassing the entire set of obstacles in contact with the dislocation at the respective time. Under these conditions, the dislocation line remains almost straight. As the obstacle strength increases, the spatial fluctuations of the obstacle density have a stronger influence on the dislocation motion and shape. If the critical angle made by dislocation segments at given obstacle at failure becomes smaller than approximately 100º, the dislocation shape becomes rather rough and loops may be left behind on the glide plane. It is difficult to capture this type of motion in analytic models.

Several studies addressed the thermally activated motion across fields of random obstacles. In this problem, each obstacle is characterized by an activation enthalpy (which may or may not be the same for all obstacles) and it is sought to determine the free enthalpy of activation for dislocation motion. In general, it turns out that the dislocation motion may indeed be characterized by an Arrhenius-type rate equation, but the activation enthalpy is a complicated function of stress, temperature and obstacle density and strength. The problem was approached analytically by Landau and Dotsenko [8] and Schlipf [9] for systems with random identical obstacles and by Arsenault and Cadman [10], Zaitsev and Nadgornyi [11] and Schoeck [12] for distributions of obstacles with two different strengths and activation enthalpies. These studies used various simplifying assumptions of which the most important is disregarding spatial correlations between obstacle failure events along given dislocation. The approach of Schoeck [12] is aimed at relaxing this limitation, but it eventually leads only to asymptotic predictions. Several computer simulations of thermally activated dislocation motion were also performed [13,14].

The effect of obstacle distribution on the strength and strain rate sensitivity (SRS) of the material was studied by Olivares and Sevillano [15] and, more recently, by Pretorius and Nembach [16]. The first group considered several obstacle distributions in the form of cells and channels representing dislocation cell structures. They concluded that the random obstacle distribution leads to the largest SRS of all cases studied. The second group presents a study of the influence of obstacle randomness on the critical resolved shear stress and claims that evidence exists that this parameter increases with increasing the degree of randomness.

In this work we revisit some of these issues in order to gain further insight into the effect of obstacle strength and spatial distribution on the SRS of the material. In particular, we are interested in the regime of large stress and large obstacle strength, for which current analytic models are thought not to apply. This is the regime in which dislocations are expected to overcome obstacles in a correlated manner.

The motivation for this analysis is provided, in part, by our recent experimental results on the SRS of dilute Al-5%Mg alloys [17]. Although the phase diagram of this alloy predicts the separation of the $Al_3Mg$ phase at temperatures below ~200°C, precipitation is sluggish and solute remains largely in solid solution. However, solute structures (clusters) are expected. It has been shown that if these structures are dissolved by a short annealing treatment before testing, the strain rate sensitivity measured at low temperatures is significantly larger (less negative) than that obtained from samples that were not annealed. The strength of the alloy is not affected by annealing. This indicates that, in addition to the nature of the obstacles, their distribution has an influence on the SRS. It should also be noted that Al-Mg alloys exhibit negative SRS at room temperature, which leads to heterogeneous deformation and reduced ductility. If dynamic strain aging (DSA) is active, the total SRS is given by the positive component associated with the interaction of dislocations with obstacles and the negative contribution from DSA. Hence, if one aims to improve the overall rate sensitivity of the material, increasing the positive component is an option that should be explored.

## 2. Model and simulation procedure

*2.1 The model*

Let us consider a random distribution of obstacles of density $\rho_0$. The density yields a characteristic length scale $l_c = 1/\sqrt{\rho_0}$. The obstacles represent small solute clusters or precipitates and $l_c$ is assumed to be much larger than the size of these entities. This is quite adequate for our purposes since the interaction with strong, long range barriers such as forest dislocations leads to the athermal component of the flow stress. The dislocation is modeled as a flexible string of line tension $\Gamma = 1/2\, Gb^2$ regardless of its type (the isotropic line tension model), where $G$ is the material shear modulus and $b$ is the Burgers vector length. Note that $l_c$ should be sufficiently large compared to the size of the obstacle for the model of the dislocation behaving as an elastic string of constant line tension to be adequate.

A dislocation segment pinned by two obstacles bows out into an arc of dimensionless radius:

$$r^* = 1/2\tau^*, \tag{1}$$

where the applied shear stress $\tau^*$ is normalized by the Orowan stress $Gb/l_c$,

$$\tau^* = \frac{1}{2r^*} = \frac{\tau \cdot l_c b}{2\Gamma} = \frac{\tau l_c}{Gb}. \tag{2}$$

The force applied by the dislocation on the respective obstacle may be computed from the line tension as

$$F = 2\Gamma \cos\left(\frac{\theta}{2}\right), \tag{3}$$

where $\theta$ is the angle made by the two branches of the dislocation impinging against the obstacle. In dimensionless form it becomes:

$$f = \frac{F}{2\Gamma} = \cos\left(\frac{\theta}{2}\right), \quad (4)$$

while the obstacle strength, $f_c$, can be expressed in a similar notation: $f_c = \cos(\theta_c/2)$.

The interaction between the dislocation and an obstacle is described in terms of an activation energy $\Delta G = 2\Gamma d \Delta G^*$, where $d$ is a characteristic interaction range which is small compared to the mean obstacle separation $l_c$. The normalized activation energy is written $\Delta G^* = \beta f_c \Delta g(f^*)$, with $f^* = f/f_c$ being the reduced dislocation-obstacle interaction force. $\Delta g$ is a monotonically decreasing function of $f^*$. It vanishes at the maximum reduced force $f^* = 1$, when the dislocation is mechanically activated, and equals one when there is no mechanical work done by the dislocation and obstacle bypassing is purely thermally activated. The function must be convex or linear for the energy barrier to be a single valued function of the perturbation (a linear function $\Delta g$ ($f^*$) corresponds to a rectangular barrier). The activation energy $\Delta G^*$ is proportional to the non-dimensional obstacle strength, $f_c$, and $\beta$ is a numerical constant.

The specific functional form of $\Delta g$ depends on the physical nature of the obstacle. This quantity can be determined from atomistic simulations of the specific interaction considered [e.g. 18]. In this study, the functional form used for the activation energy in [8] was selected:

$$\Delta g = \left(1 - f^*\right)^2. \quad (5)$$

This allows us to compare the results with the analytic model presented in this reference.

The probability that the dislocation overcomes an obstacle is given by the usual Arrhenius form:

$$p = \exp\left(-\frac{2\Gamma d \Delta G^*}{kT}\right) = \exp(-\alpha \Delta G^*), \quad (6)$$

where the dimensionless temperature $\frac{1}{\alpha} = \frac{kT}{2\Gamma d}$ is used. Since the effect of line tension $\Gamma$ and the characteristic interaction distance $d$ are lumped into parameter $\alpha$, increasing the temperature is equivalent to decreasing the line tension $\Gamma$ or the interaction distance $d$.

The strain rate sensitivity parameter, $m$, is evaluated based on the computed dislocation velocity, $v$, under given stress as

$$m = \frac{\partial \log \tau^*}{\partial \log v^*}, \quad (7)$$

where the velocity is normalized as $v^* = v/l_c \varsigma$, with $\varsigma$ being the attempt frequency for overcoming obstacles ($1/\varsigma$ is the only implicit time of the problem). Inertia was not considered and the transition time between consecutive stable dislocation configurations was neglected compared to the waiting time spent at each stable configuration. Note that the activation volume of the dislocation glide process is given by:

$$V = kT \frac{\partial \log v}{\partial \tau} = \frac{kT}{m\tau} \propto \frac{1}{m\alpha\tau}. \quad (8)$$

*2.2 The simulation procedure*

The "circle-rolling procedure" devised by Foreman and Makin [19] was modified to implement thermally activated dislocation motion. In each simulation, arrays of approximately 10,000 obstacles were generated and a single dislocation was moved across.

The model is large enough to prevent strong effects due to the periodic boundary conditions used in the direction perpendicular to the glide direction [6,7]. All obstacles have the same strength and activation enthalpy.

At given stress, each dislocation segment bows out until it meets a new obstacle, equilibrium at one of the two pinning obstacles is lost, or the equilibrium radius $r^*$ is reached. A mechanically stable configuration is obtained by verifying these conditions for all obstacles currently in contact with the dislocation, until every dislocation segment reaches equilibrium. The critical resolved shear stress may be identified as the minimum stress at which no stable configuration can be found.

Once a stable configuration is identified under the applied stress $\tau^*$, the time is incremented by a small increment $dt^* = \varsigma \cdot dt$. For every time increment, a random number generator is used to mimic the local activation process according to equation (6) for all obstacles in contact with the dislocation. The dislocation is advanced to the next stable configuration by releasing all activated obstacles. This process is repeated and the glide distance $l^* = l/l_c$ during the total simulation time $t^*$ is measured. The non-dimensional dislocation velocity is computed as $v^* = l^*/t^*$. The strain rate sensitivity parameter is evaluated using eqn. (7) and by subjecting the sample in separate simulations to shear stresses $\tau^*$ and $\tau^* + d\tau^*$, with $d\tau^*$ on the order of $0.01\,\tau_c^*$, where $\tau_c^*$ is the athermal critical resolved shear stress. Note that this procedure allows for multiple obstacles to be activated in the same time interval $d\tau^*$ and hence correlated activation events are permitted.

In arrays with large obstacle strength, dislocation loops are left behind in the glide plane. The effect of the loops is not considered as the array is continuously regenerated in front of a moving dislocation. The field-mediated interaction of dislocation segments is neglected, and the mechanics is controlled by the line tension only.

A large number of replicas of same obstacle density ($\rho_0$), strength ($f_c$) and distribution were generated by using different random seeds and the results were obtained by averaging over all these realizations.

In the computer code, all obstacles have a user defined data structure which includes the obstacle coordinates, strength and current state (bypassed by the dislocation, in contact or not-yet-in-contact with the dislocation). The dislocation is represented by a two-way linked list. Each node in the list stands for an obstacle in contact with the dislocation. Information such as the segment length and angle relative to the glide direction, pointers to previous and next obstacles is also included. This information is dynamically updated as the dislocation moves through the array.

*2.3 Generation of non-random obstacle arrays*

Generating obstacle arrays with pre-defined obstacle distribution is not trivial. The distribution is specified by imposing the radial pair distribution function, *g(r)*. This function represents the density of obstacles in an annulus of thickness *dr* located at distance *r* from a specified obstacle, normalized by the mean obstacle density. For proper comparison of the results, the mean density $\rho_0 = 1/l_c^2$ was kept constant in all simulations.

To generate the obstacle array, a procedure commonly used to derive potentials of mean force from known radial distribution functions was used. It is based on the iterative

Boltzmann inversion method [e.g. 20]. If one desires to develop an atomistic model with desired pair distribution function, $g_{target}(r)$, a potential of mean force is inferred as

$$U_0(r) = -kT \log[g_{target}(r)]. \qquad (9)$$

This potential is then used to perform a molecular dynamics simulation in 2D. After equilibration, this produces a pair distribution function $g_1(r)$, which is different from the target radial distribution function $g_{target}(r)$. The potential is then corrected and the procedure is repeated until the convergence of the distribution function to the targeted one. The correction of the potential in the *n*-th iteration (based on $g_n(r)$) is given by:

$$U_{n+1}(r) = U_n(r) + kT \log\left[\frac{g_n(r)}{g_{target}(r)}\right], \qquad (10)$$

If the target radial distribution function has a simple, realistic function form, the procedure converges in several iterations.

Once the potential is determined, many sets of obstacles with desired $g_{target}(r)$ may be generated by simply running molecular dynamics in 2D and saving configurations. The "atom" positions define the obstacle array.

## 3. Results and discussion

Results for the random and correlated obstacle distributions are presented in sub-sections 3.1 and 3.2, respectively. The normalized SRS variation with the obstacle strength, $f_c$, the temperature, $\alpha$, and the normalized applied stress, $\tau^*/\tau_c^*$ is studied. A simple model that assists data interpretation is presented and comparisons are made with relevant experimental and theoretical results.

*3.1 Random obstacle distributions*

The critical resolved shear stress $\tau_c^*$ at which dislocations glide through in absence of thermal activation has been evaluated theoretically and numerically [2,3,4,19]. A square lattice distribution of obstacles gives, $\tau_c^* = f_c$. Friedel's result for a random distribution of point obstacles is $\tau_c^* = f_c^{3/2}$, which holds for weak obstacles of strengths below approximately $f_c = 0.5$. At larger $f_c$, Friedel's relation overestimates the critical stress. This error is encountered in all theoretical derivations that make the assumption that obstacle failure events are spatially uncorrelated. The results of the present analysis are shown in Fig. 1 along with computer simulation results for the same system by Foreman and Makin [19]. These critical stresses are used below as reference.

The variation of the inverse SRS parameter, $1/m\alpha$, with the obstacle strength is shown in Fig. 2. The data are obtained for two applied stresses, $\tau^* = 0.5\,\tau_c^*$ and $0.9\,\tau_c^*$ to represent small and large stress conditions, respectively. Here, $\tau^*$ is different for the various points along the curves, since $\tau_c^*$ is a function of $f_c$. It results that at low stress, $1/m\alpha$ is proportional to $f_c$, i.e. the SRS parameter $m$ is proportional to the temperature and inversely proportional to the obstacle strength. The same relationship was found in [7,13,15] and seems to agree with experimental data [21].

When the applied stress is large and for obstacles with $f_c < 0.6$, $1/m\alpha$ does not follow the same scaling, rather, $m$ is independent of the obstacle strength $f_c$ in this range. When $f_c > 0.6$, the linear scaling of $1/m\alpha$ with $f_c$ is recovered. To understand this behavior, let us consider an Arrhenius relation for the non-dimensional dislocation velocity [8]:

$$v^* = \Omega(\tau^*)\exp(-\alpha\Delta G^*(f^*)). \tag{11}$$

This expression is postulated based on previous theoretical work and numerical results [7,13]. It was proven only in the context of weak obstacles, situation in which the assumption of uncorrelated obstacle failure holds. Then, it can be shown that the activation enthalpy for the entire dislocation is identical to that for overcoming an individual obstacle. $\Omega$ represents the average distance (normalized by the mean obstacle separation $l_c$) a dislocation advances from a stable configuration to the next as a result of a single activation event.

Since under these conditions the dislocation line is almost straight, the force acting on an obstacle may be expressed in terms of the applied shear stress as:

$$f^* = \frac{l^*\tau^*}{f_c} \approx \frac{(\tau^*)^{1-n}}{f_c}. \tag{12}$$

The second part of the equation results by using the expression $l^* \approx \kappa'(\tau^*)^{-n}$ relating the distance between obstacles along the dislocation line to the applied stress [1,3,4], ($\kappa' \approx 1$). This expression is valid under the same assumptions as Friedel's relation, but it fits numerical results over a broad range of obstacle strengths and applied stress. The exponent $n$ = 1/3 for randomly distributed obstacles and $n = 0$ for the square lattice distribution ($l^* = 1$).

Substituting eqn. (11) in (7) and using eqn. (12) one obtains:

$$\frac{1}{m} = \frac{\partial \log \Omega}{\partial \log \tau^*} - \alpha(1-n)\frac{d\Delta G^*}{d \ln f^*}. \tag{13}$$

With the normalized activation energy given by eqn. (5), the inverse SRS parameter results:

$$\frac{1}{m} = \frac{\partial \log \Omega}{\partial \log \tau^*} + 2\alpha(1-n)\beta f_c f^*(1-f^*) \tag{14}$$

The reciprocal strain rate sensitivity has two components: one defined by the stress dependence of $\Omega$, and the other proportional to the obstacle strength and the normalized inverse temperature, $\alpha$. Referring to the data shown in Fig. 2, it results that at low applied stress ($\tau^* = 0.5\tau_c^*$), the first term has no contribution to $1/m$ ($\Omega$ is a constant proportional to the average obstacle spacing, $l_c$) and $1/m\alpha$ results proportional to $f_c$. This is also supported by the fact that the two curves corresponding to different $\alpha$ collapse. Interestingly, although the assumptions used to derive eqn. (14) are rather restrictive and considered to hold only for weak obstacles, the model fits well the numerical result even for large $f_c$. The interpretation of this observation will be discussed later in this Section.

At large applied stress and low obstacle strength, $1/m\alpha$ is independent of $f_c$ and proportional to $1/\alpha$ which indicates that the second term in eqn. (14) contributes little to SRS. This is interpreted as a manifestation of the stress dependence of $\Omega$. As $f_c$ increases, $\Omega$ becomes stress-independent and the second term dominates just as in the case of low applied stress. The slope turns out to be smaller than that obtained for $\tau^* = 0.5\tau_c^*$, which can be also justified based on the second term of eqns. (14) (by using eqn. (12) and the numerical relationship between $\tau_c^*$ and $f_c$ shown in Fig. 1).

To gain insight into the origin of these dependencies it is useful to look at the mechanism by which the dislocation moves. At low stresses the dislocation moves by unzipping. When the stress increases, the motion becomes jerky, i.e. the dislocation advances through a series of large jumps and obstacles are overcome in a correlated manner. The magnitude of the jumps depends on the applied stress and is not determined by the length scale $l_c$. A similar result was obtained by Mohles and Ronnpagel [22] when using a stress of $\tau^* = 0.9\tau_c^*$. This transition may be observed in animations of the simulated glide process. Jerky motion of dislocations was also reported in some models of thermally activated dislocation glide [23,24] and is commonly observed by electron microscopy.

The critical stress at which this transition occurs requires some discussion. Let us consider a dislocation segment pinned at O, P and Q (Fig. 3). If the dislocation moves by unzipping, as the segment is released from P, a new equilibrium configuration OP'Q is reached; bypassing obstacles O and Q is independent of the event at P. Then, $\Omega$ is proportional to the mean area per obstacle ($1/\rho_0$) and is independent of the applied stress. As the applied stress increases, a critical stress is reached (denoted by $\bar{\eta}\tau_c^*$, $\bar{\eta} < 1$) at which this mechanism ceases to be valid. To determine this threshold, the condition that release at P implies release at O or/and Q is written (the condition for correlated obstacle release). The limit situation is that in which Q is released when the dislocation reaches configuration OP'Q. The condition of critical state at Q reads:

$$\cos\left[\frac{1}{4}(\pi-\gamma)+\frac{1}{2}\left(a\cos\left(2\bar{\eta}\tau_c^* l^* \sin\left(\frac{\gamma}{2}\right)\right)+a\cos\left(\bar{\eta}\tau_c^* l^*\right)\right)\right] = f_c, \qquad (15)$$

which provides an equation for $\bar{\eta}$ in terms of the obstacle strength $f_c$. $\tau_c^*$ and $\gamma$ are themselves functions of $f_c$. $\gamma$ has a narrow distribution in the neighborhood of $\pi$ for weak obstacle, while for strong obstacles the width of the distribution increases. As we are considering the limit of unzipping, the expression $l^* \approx \kappa'(\tau^*)^{-n} = \kappa'(\bar{\eta}\tau_c^*)^{-n}$ can be used. Also, the distribution of $\gamma$ is replaced by its mean, i.e. $\gamma = \pi$. Equation (15) leads to $\bar{\eta} = 0.57$ for $f_c = 0.1$. As $f_c$ increases, $\bar{\eta}$ increases monotonically such that for $f_c = 1$, $\bar{\eta} \approx 1$. Hence, obstacles of large strength stabilize the unzipping mechanism. Of course, obstacles of strength $f_c \sim 1$ are probably too strong to represent realistic situations; this limit is considered in this work in order to have a complete physical picture.

Returning to the discussion of the data in Fig. 2, it results that the validity of eqn. (14) and the underlying assumptions is not conditioned by the obstacle strength per se; rather, it is associated with the unzipping mechanism (uncorrelated obstacle bypassing).

The variation of $1/m\alpha$ with temperature ($1/\alpha$) is shown in Fig. 4. As discussed, at low stress (creep conditions), the second term in eqn. (14) dominates and $1/m\alpha$ is independent of temperature [7,13]. At large stresses, $1/m\alpha$ (and the activation volume) is approximately proportional to the temperature. This is in general agreement with experimental results [e.g. 25].

Let us consider now the effect of stress on the SRS parameter. Figure 5 shows simulation results for $1/m\alpha$ versus the normalized applied shear stress, $\eta = \tau^*/\tau_c^*$ for random arrays with $f_c = 0.1$, 0.5 and 1.0 representing weak, intermediate and strong obstacles. To assist data interpretation, let us consider eqn. (14) with $f^*$ given by eqn. (12). At low stress (creep conditions), when $\Omega$ is essentially independent of stress, this leads to

$$\frac{1}{m\alpha} = 2(1-n)\beta f_c \frac{\left(\eta \tau_c^*\right)^{1-n}}{f_c}\left[1 - \frac{\left(\eta \tau_c^*\right)^{1-n}}{f_c}\right]. \tag{16}$$

This equation can be used for random distributions ($n = 1/3$) and for the square lattice arrangement ($n = 0$) and leads to very good agreement with the numerical data for $\eta < \bar{\eta}$ (shown by thick solid and dashed lines in Fig. 5, for $n = 1/3$ and $n = 0$, respectively). The equation was derived under restrictive conditions: no spatial and temporal correlations between obstacle release events, the dislocation line is almost straight and the relation $l^* \approx \kappa'(\tau^*)^{-n}$ holds. These conditions have been associated in the literature with weak obstacle. The results presented in Figs. 2 and 5 indicate that the model can be used for strong obstacles too, provided the dislocation moves by unzipping ($\eta < \bar{\eta}$).

The predictions of the formulation developed by Landau and Dotsenko [8] are also shown in Fig. 5a (valid for weak obstacles only) by thin solid and dash lines corresponding to $\alpha = 80$ and 30, respectively.

For $\eta < \bar{\eta}$, $1/m\alpha$ is independent of temperature and hence the curves corresponding to simulations performed with different $\alpha$ superimpose. It is also interesting to note that the Cottrell-Stokes law, i.e. the stress-independence of $1/m\alpha$, is not exactly fulfilled for any obstacle strength and applied stress. However, for weak obstacles and low levels of stress ($\eta < \bar{\eta}$), $1/m\alpha$ depends weakly on stress ($\eta$).

At large stresses, $\eta > \bar{\eta}$, the predictions of eqn. (16) remain a good match to the numerical results for the square lattice distribution, but do not follow the data for the random obstacle distributions. As discussed, this is due to the fact that $\Omega$ becomes stress-dependent and the dislocation motion becomes jerky (correlated obstacle bypassing). The SRS parameter $m$ decreases very fast with increasing applied stress and becomes dependent on temperature. Increasing the temperature leads to a more pronounced reduction of $m$. The critical stress $\bar{\eta}$ at which the upturn of $1/m\alpha$ is observed depends strongly on the obstacle strength (eqn. (15)). For the strongest obstacles ($f_c = 1$), $\Omega$ is independent of stress and the dislocation moves by unzipping at all stresses ($\bar{\eta} \approx 1$); the temperature has no influence on the plot (*m* varies as $1/\alpha$).

These results indicate that the mechanism by which the dislocation moves (correlated vs. uncorrelated obstacle bypassing) is controlled by both the applied stress and the obstacle strength. In fields of weak obstacles, correlated obstacle bypassing (jerky motion) associated with reduced SRS is observed at relatively small stresses. This phenomenon is not controlled by the nature of the obstacles as long as their size *d* is sufficiently small compared to the inter-obstacle spacing.

Figure 5 also indicates that the SRS parameter *m* is larger when the obstacles are weaker. The random distribution leads to larger *m* than the square lattice distribution of same obstacle density. This difference is more pronounced for strong obstacles. In [15] it was also reported that the random distribution leads to the largest SRS parameter out of several random and non-random obstacle distributions considered. This issue is revisited in Section 3.2.

In experiments it is observed that the activation volume (eqn. (8)) decreases continuously with stress [25,26]. The data in [26] can be mapped to a curve with a weak maximum at intermediate stresses. Although in agreement with the present results, a quantitative comparison with the experiments is not possible due to the simplicity of the model.

The transition from smooth to jerky dislocation motion at $\eta = \bar{\eta}$ is not reported, to our knowledge, in previous works. Altintas *et al*. [13] discuss that the deformation is markedly inhomogeneous at low temperatures and becomes homogeneous at low stresses as the temperature increases, while at high stresses, the deformation remains inhomogeneous at all temperatures. In [13], a set of parallel independent glide planes are considered and what they refer to as heterogeneous deformation is caused by variability from plane to plane, i.e. from realization to realization of the array of obstacles.

**3.2 Correlated obstacle distributions**

Let us consider next distributions of obstacles characterized by additional (larger) length scales. The fundamental length scale associated with the mean obstacle density, $l_c = 1/\sqrt{\rho_0}$, is kept constant and longer wavelength density fluctuations are introduced.

To demonstrate this effect, two types of distributions are considered, corresponding to two imposed pair distribution functions, g(r). These are shown in Fig. 6 and are denoted as configurations 1 and 2. The figure also shows g(r) for the random obstacle distribution, which is a constant of *r*. In configuration (1), the obstacles are clustered, with an additional length scale defining the mean distance between clusters of about $3l_c$. No long range order exists beyond this scale. Configuration (2) is approaching the long range order of the 2D closest packed structure (the triangular lattice), with little or no clustering.

The critical resolved shear stress ($\tau_c^*$) for the two distributions is shown in Fig. 7 as a function of the obstacle strength, $f_c$. The curve corresponding to the random distribution (Fig. 1) is also shown. For weak obstacles, $f_c \leq 0.5$, all three distributions give the same critical shear stress. Clusters of weak obstacles in configuration 1 tend to be stronger than individual obstacles due to the small obstacle separation within each cluster. At the same time, the mean distance between clusters is much larger than $l_c$, which should lead to a smaller $\tau_c^*$. These two effects balance each other in the case of weak obstacles. In the case of strong obstacles, the second effect dominates in configuration 1. $\tau_c^*$ for configuration 2 is identical to that of the random lattice for all obstacle strengths.

The strain rate sensitivity analysis of the two distributions is shown in Fig. 8 for weak ($f_c = 0.1$) and strong ($f_c = 1$) obstacles. Interestingly, for weak obstacles the SRS parameter is identical for the two configurations and identical to that of the random distribution. This is also observed in arrays of strong obstacles ($f_c = 1$), for which all curves are similar at low stress. However, important differences are observed in presence of stronger obstacles at large stresses. Under these circumstances, the additional length scales associated with the obstacle distribution become important. The threshold ratio $\bar{\eta}$ appears to be sensitive to the nature of the distribution. For configuration 1 $\bar{\eta} \approx 0.5$, for configuration 2, $\bar{\eta} \approx 0.8$, while for the random distribution $\bar{\eta} \approx 1$.

The presence of obstacle clusters leads to substantially smaller $m$ at large stresses. Furthermore, configuration 2 leads to a SRS parameter $m$ larger than that obtained for random distributions.

These results show that under creep conditions the SRS is only weakly influenced by the distribution of obstacles. At large stresses, the obstacle distribution becomes important. This is actually expected since at large stresses the dislocation bypassed obstacles in a correlated manner (jerky motion) and its motion should be influenced by how obstacles are distributed. These observations also indicate that characterizing the strength and SRS by a single parameter, i.e. the mean obstacle density, is insufficient and that higher moments of the obstacle density distribution must be also considered.

## 4. Conclusions

The study presented in this article shows that the analysis of the effect of obstacles on the critical resolved shear stress and strain rate sensitivity can be divided in two parts function of the mechanism by which dislocations move. If the motion is smooth and takes place by unzipping, the strain rate sensitivity is proportional to the temperature and inversely proportional to the obstacle strength. Its variation with the applied stress is predicted by a simple model which also captures several other features of the overall behavior. A threshold stress is identified (and predicted by the model) at which a transition to jerky dislocation motion is observed. During the jerky motion the dislocation bypasses obstacles in a correlated manner. The transition stress increases with the obstacle strength indicating that strong obstacles stabilize the unzipping mode. If the dislocation moves in a jerky fashion, the strain rate sensitivity decreases and becomes rather insensitive to the obstacle strength.

The strain rate sensitivity is greatly affected by spatial correlations introduced in the distribution of obstacles, especially when the dislocation motion is jerky. This parameter is affected more than the critical resolved shear stress. Hence, models aimed at predicting these global measures should account for higher moments of the obstacle density distribution, in addition to the mean density.

**Acknowledgment**

This work was supported, in part, by the NSF through grant CMS-0502891.

**Figure captions**

Figure 1. Normalized critical resolved shear stress $\tau_c^*$ in absence of thermal activation, function of the obstacle strength, $f_c$. The results from the present simulations are shown together with numerical results from Ref. [19] and with predictions of Friedel's model.

Figure 2. Variation of the inverse SRS parameter with the obstacle strength at small and large applied stress and at two different normalized temperatures.

Figure 3. Schematic representation of two successive dislocation configurations (shown by solid and dash lines) used to derive the critical stress $\bar{\eta}\tau_c^*$ at which the dislocation motion mechanism changes from unzipping to jerky.

Figure 4. Variation of the normalized SRS $1/m\alpha$ with the temperature ($1/\alpha$) for arrays of weak and strong obstacles subjected to large stress.

Figure 5. Variation of the normalized SRS parameter with stress in random (circles) and square lattice-type arrays (squares) of obstacles with a) $f_c = 0.1$, b) $f_c = 0.5$ and c) $f_c = 1$. The thick curves represent predictions of eqn. (16) for the two types of arrays (solid line for the random arrays). The thin solid and dash lines in a) are predictions of the model in [8] for the two extreme temperatures considered here, $\alpha = 30$ and 80, respectively. The dotted lines in a) are simply guide to the eye for the data points corresponding to the random array in the regime $\eta > \bar{\eta}$.

Figure 6. Pair distribution functions for two obstacle distributions considered in this study.

Figure 7. Normalized critical resolved shear stress $\tau_c^*$ in absence of thermal activation, function of the obstacle strength, $f_c$, for the random array and for the two distributions shown in Fig. 6.

Figure 8. Variation of the normalized SRS parameter with stress for random arrays (circles) and for the two configurations shown in Fig. 6, with a) $f_c = 0.1$ and b) $f_c = 1$. The curve represents the predictions of eqn. (16) for the random array.

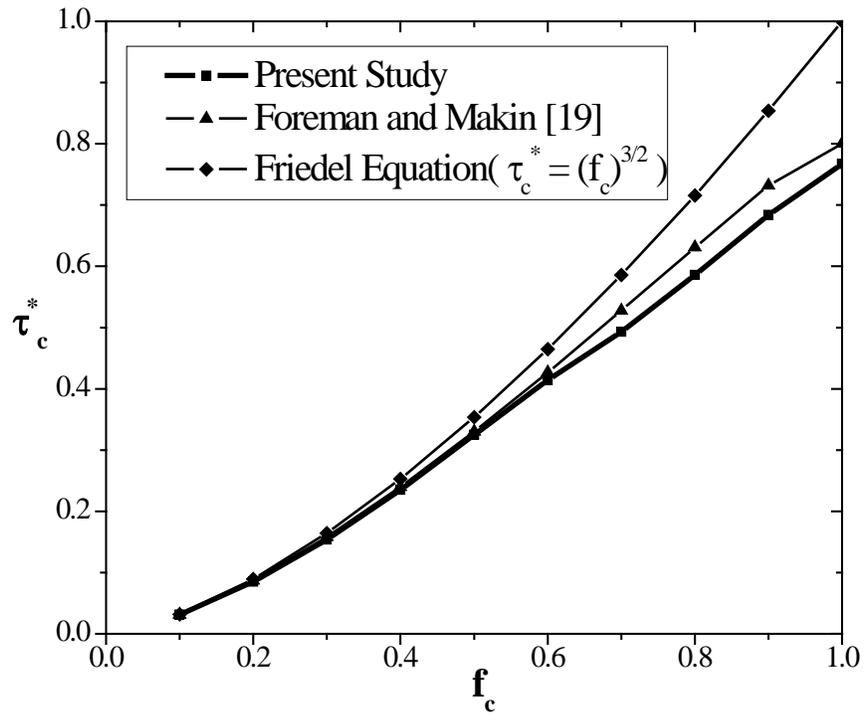

Figure 1

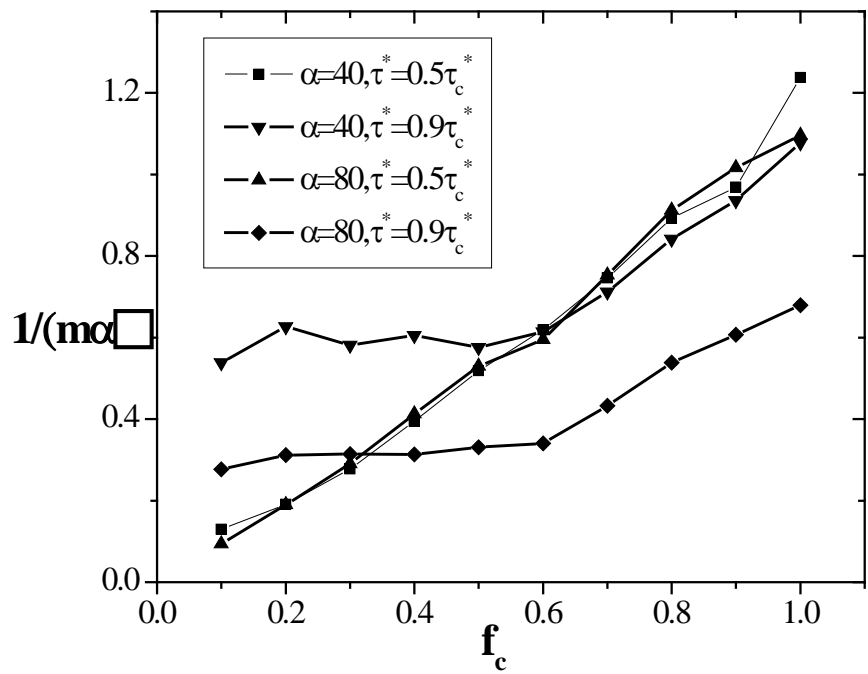

Figure 2

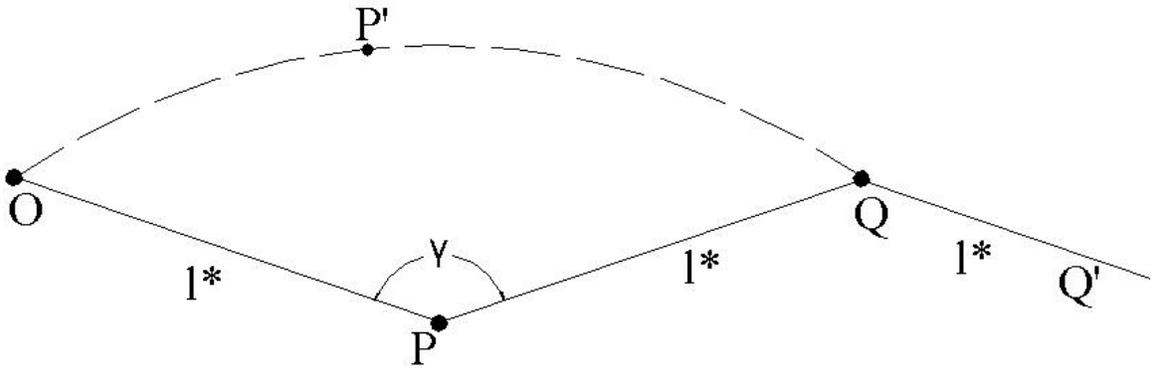

Figure 3

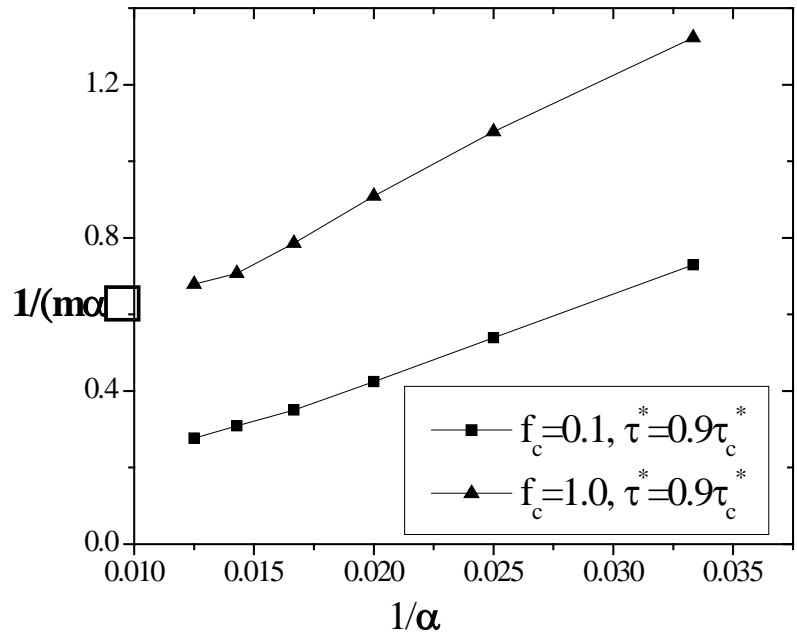

Figure 4

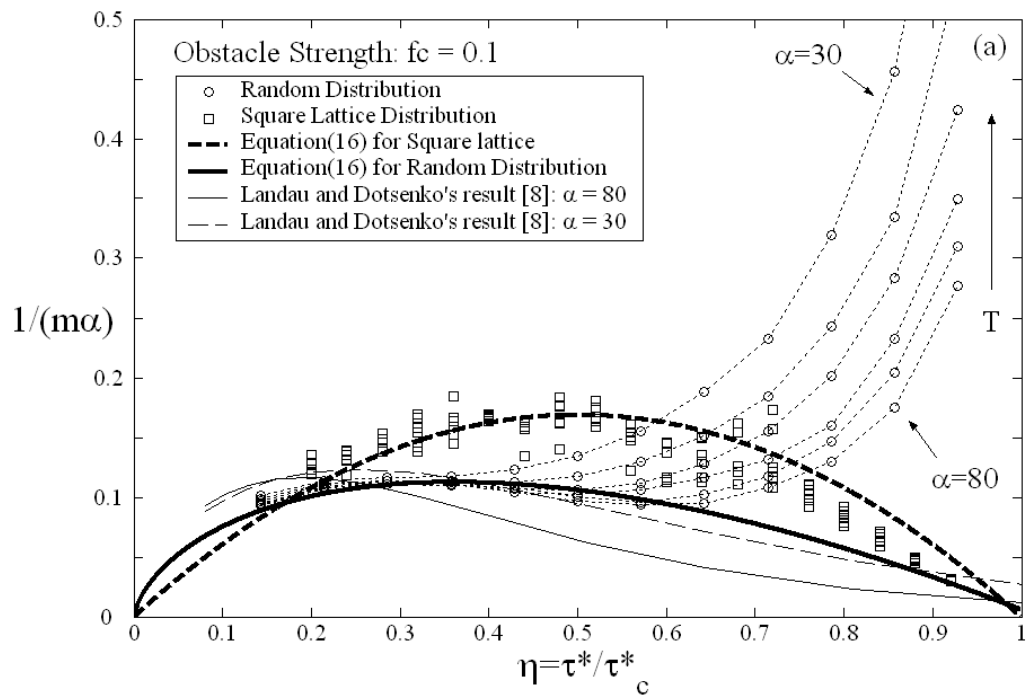

Figure 5a

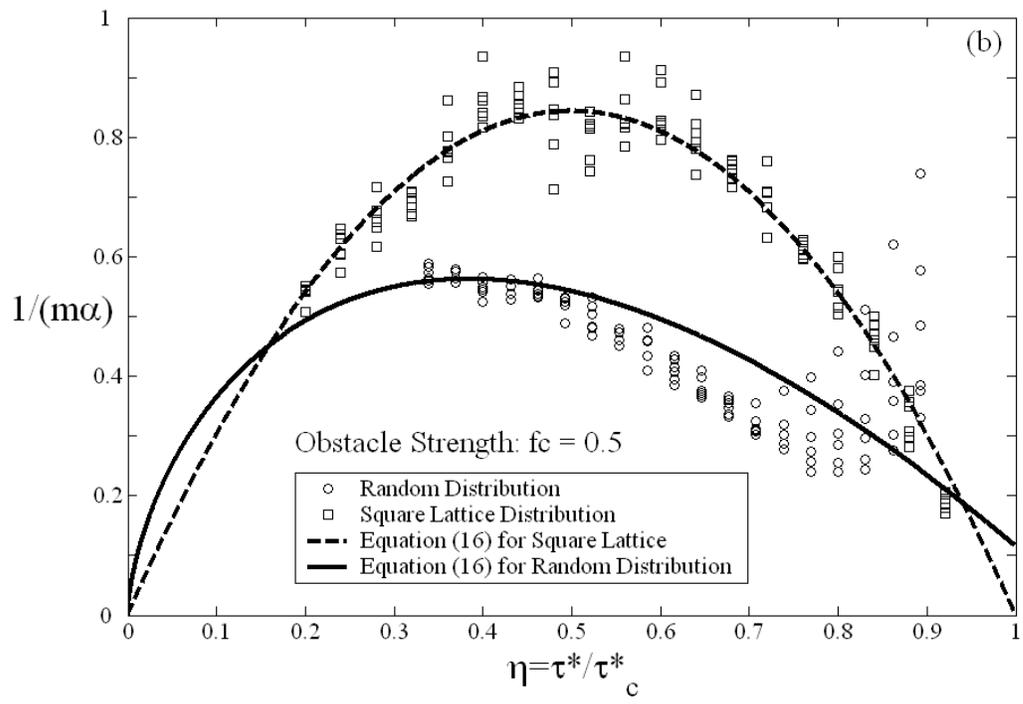

Figure 5b

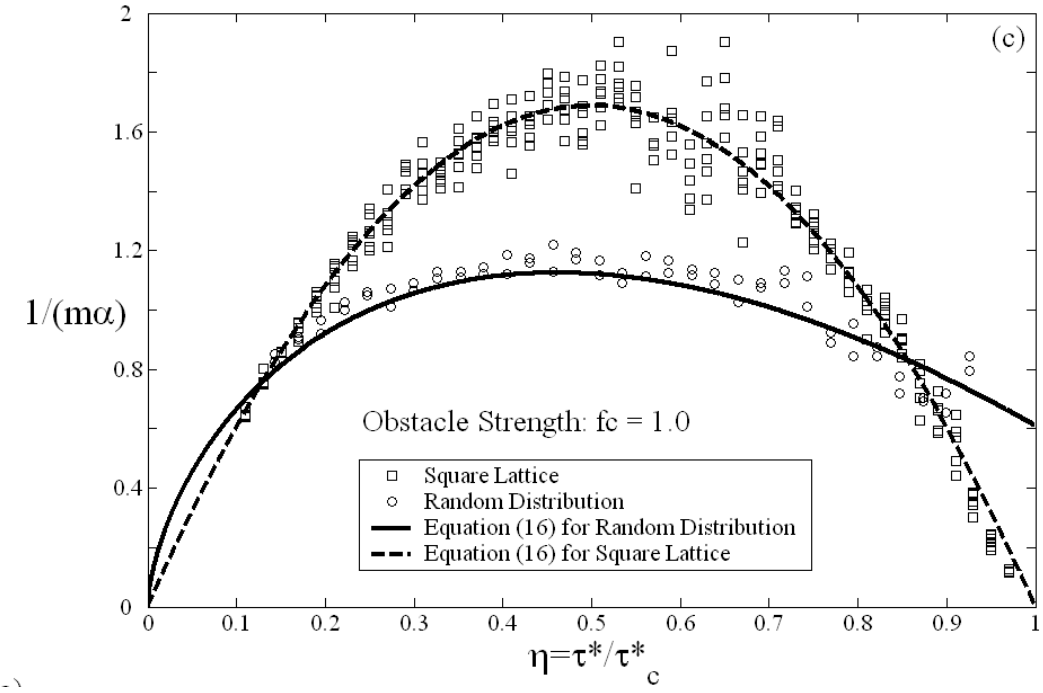

Figure 5c

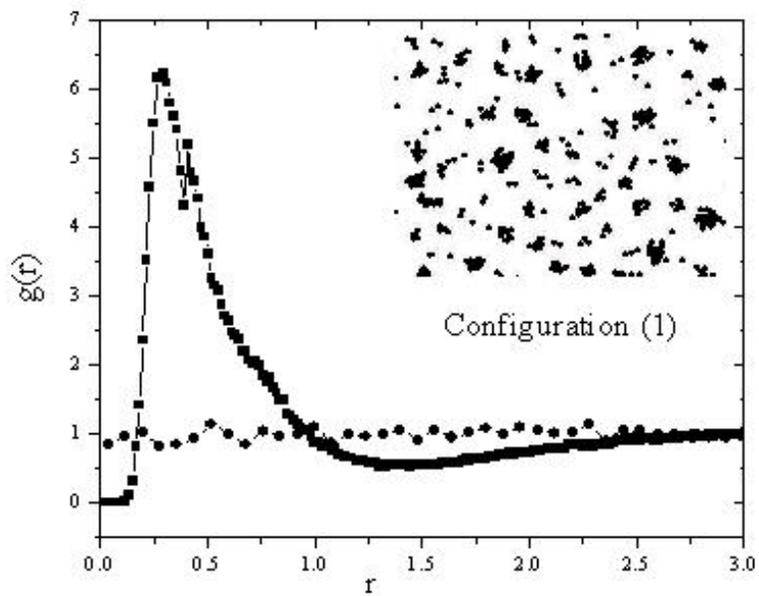

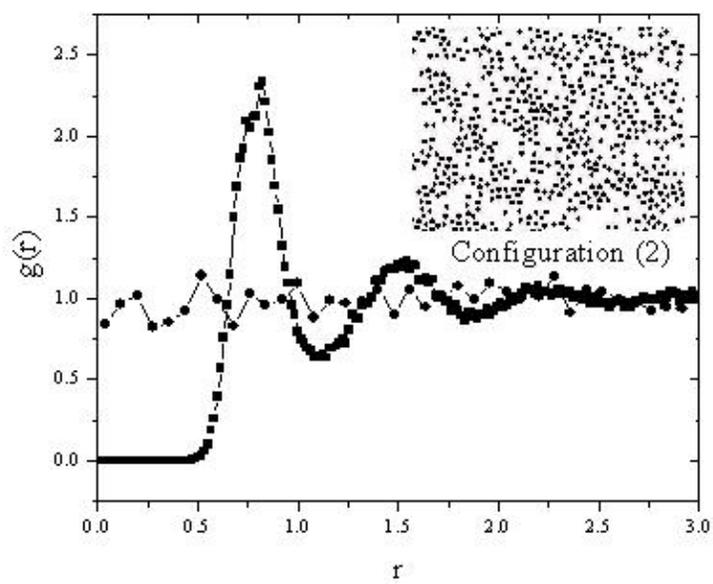

Figure 6

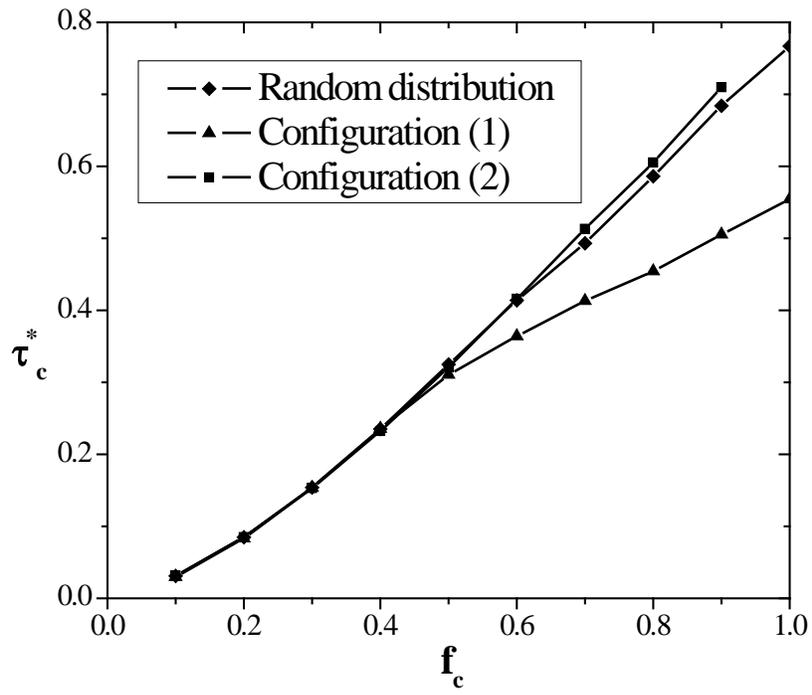

Figure 7

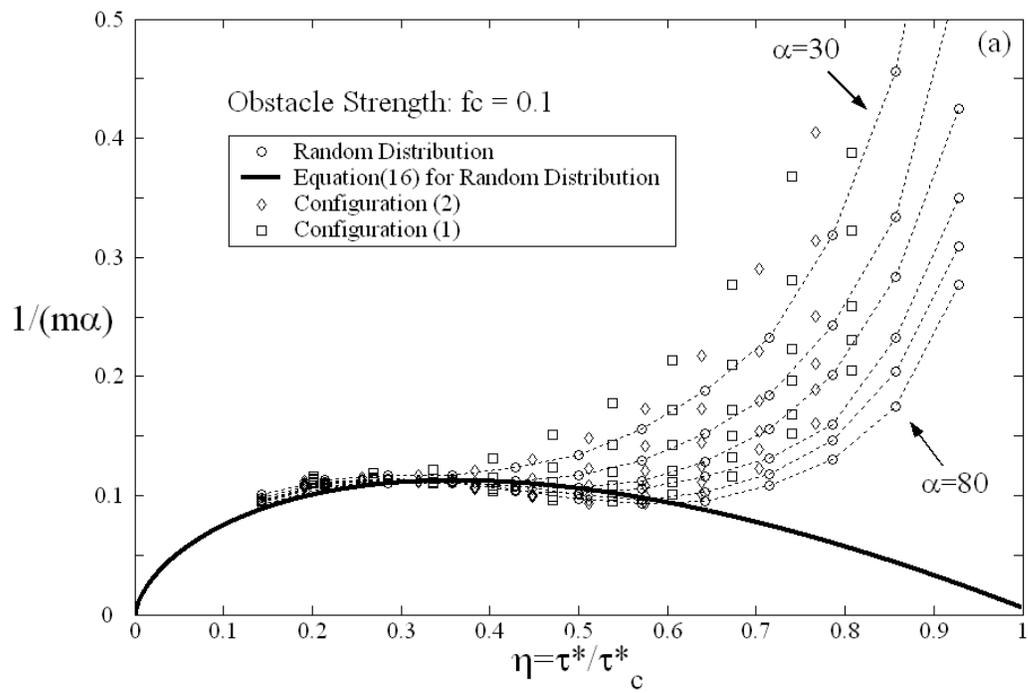

Figure 8a

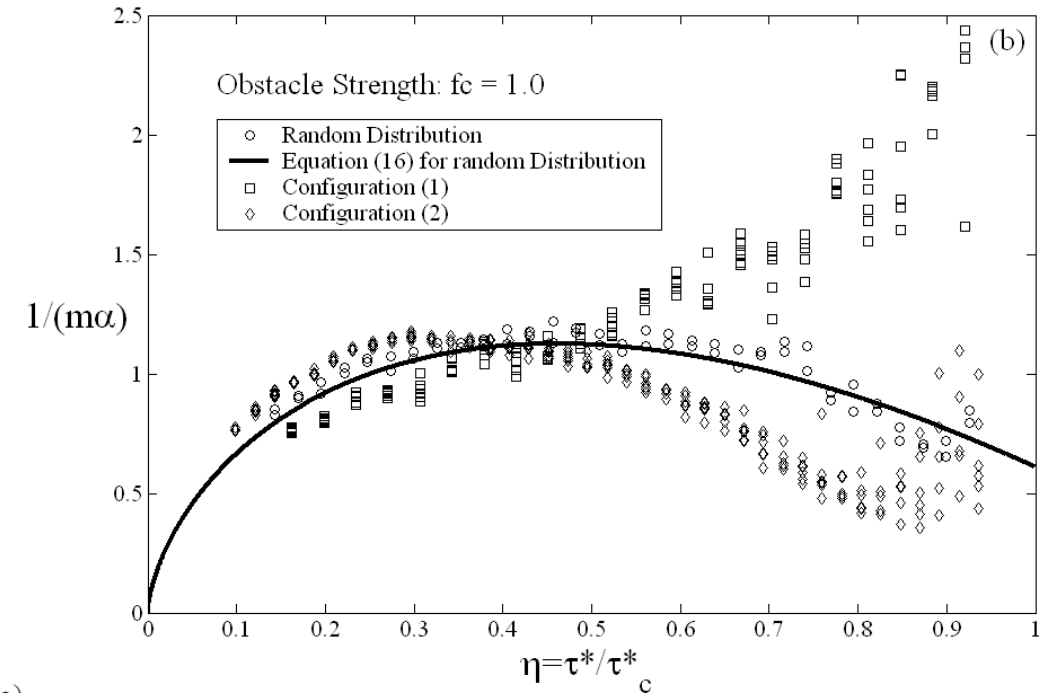

Figure 8b